\documentclass[12pt]{article}
\usepackage{amsmath,amssymb,graphicx,mathrsfs,hyperref,slashed,setspace}

\usepackage{float}
\usepackage{latexsym}
\usepackage{bm}
\usepackage{blindtext}
\hypersetup{
	colorlinks=true,
	linkcolor=blue,
	filecolor=black,
	urlcolor=black,
	citecolor=blue,
}

\newcommand{\hide}[1]{}

\newcommand{\be}{\begin{equation}}
\newcommand{\ee}{\end{equation}}
\newcommand{\bea}{\begin{eqnarray}}
\newcommand{\eea}{\end{eqnarray}}

\def\({\left(} \def\){\right)}

\begin{document}
\title{\vspace{-1.8in}
{Black holes as frozen stars}}
\author{\large Ram Brustein${}^{(1)}$,  A.J.M. Medved${}^{(2,3)}$,
Tamar Simhon${}^{(1)}$
\\
\vspace{-.5in} \hspace{-1.5in} \vbox{
\begin{flushleft}
 $^{\textrm{\normalsize
(1)\ Department of Physics, Ben-Gurion University,
   Beer-Sheva 84105, Israel}}$
$^{\textrm{\normalsize (2)\ Department of Physics \& Electronics, Rhodes University,
 Grahamstown 6140, South Africa}}$
$^{\textrm{\normalsize (3)\ National Institute for Theoretical Physics (NITheP), Western Cape 7602,
South Africa}}$
\\ \small \hspace{1.07in}
   ramyb@bgu.ac.il,\  j.medved@ru.ac.za,\ simhot@post.bgu.ac.il
\end{flushleft}
}}
\date{}
\maketitle
\begin{abstract}
We have recently proposed a model for a regular black hole, or an ultra-compact
object, that is premised on having maximally negative radial pressure throughout the  entirety of the object's interior. This model can be viewed as that of a highly entropic configuration of fundamental, closed strings near the Hagedorn temperature, but from the
perspective of an observer who is ignorant about the role of quantum physics
in counteracting against gravitational collapse. The advantage of this
classical perspective is that one can use Einstein's equations to define a classical geometry and investigate its stability. Here, we complete the model by studying  an important aspect of this framework that has so far been overlooked: The geometry and composition of the outermost layer of the ultra-compact object, which interpolates between the bulk geometry of the object and the standard Schwarzschild vacuum solution in its exterior region. By imposing  a well-defined set of matching conditions, we find a  metric that describes this transitional layer and show that it satisfies all the basic requirements; including the stability of the object when subjected
to small perturbations about the background solution.  In fact, we are able
to show  that, at linearized order,  all geometrical and matter fluctuations are perfectly
frozen in the transitional layer, just as they are known to be in the bulk of the object's interior.
\end{abstract}
\maketitle

\newpage

\section{Introduction}

The final state of matter collapsing under its own gravity, what is now universally known as a
black hole (BH), had been initially termed as a ``frozen star'' \cite{Ruffini:1971bza}. From the perspective of an outside observer, the collapse continues, formally, for an infinite period of time
and deviations from the static Schwarzschild geometry decay exponentially in time.   The scale for
the latter  is the light-crossing time across the collapsing star, and having  such a short time scale
for perturbations to decay  means that the star is essentially  ``frozen".

The rebranding of frozen stars as BHs came about once the singular nature of their classical solutions was finally confirmed.  The first efforts along this line
can be traced to the likes of  Raychaudhuri \cite{ray} and Komar \cite{komar}.  A further sign that something was amiss can be seen
in the works of Buchdahl \cite{Buchdahl}, Chandrasekhar \cite{chand1,chand2} and  Bondi \cite{bondi}, who used the formalism of  general relativity  to show that ``normal" matter cannot be stable when confined to a small-enough radius.  Penrose and Hawking formalized these ideas in mathematical terms with their singularity theorems  \cite{PenHawk1,PenHawk2}.

As singularities are untenable in the quantum realm because of the associated violations of unitarity, a common assumption is  that regularity will be preserved once  quantum mechanics is correctly incorporated.  A popular expectation is that quantum effects at the Planck scale will resolve the BH singularity and  replace infinities with large-but-finite densities.  However, attempts at realizing this expectation have so far failed to succeed. As it turns out,  if the resolution scale is much smaller than the Schwarzschild scale, then the quantum emission of particles from the object will be such that the emitted energy greatly exceeds the original BH mass \cite{frolov,visser}. Our conclusion, therefore, is that deviations from general relativity that extend over horizon-sized scales should be a minimal requirement for a  non-singular BH.

Ultra-compact objects (UCOs)  have become shorthand for extremely dense astrophysical objects
that are regular but otherwise behave ---  for the most part --- like the
BHs of general relativity along with its standard semiclassical
extensions. (See \cite{carded} for an exhaustive review and
``status report''.)
The collapsed polymer model of a BH \cite{strungout} was born out of the idea that
a viable UCO  would be one that contains a maximally entropic fluid  throughout its interior,  which in turn means that it is described by a highly quantum state of exotic matter \cite{inny}.
With inspiration from various sources \cite{AW,SS,LT,HP,DV}, what was eventually proposed was a trans-Hagedorn configuration of long, closed and interacting strings. One finds that this polymer model can explain  all known properties of Schwarzschild
BHs \cite{emerge}. Given a significant enough departure away from equilibrium \cite{ridethewave}, this description can also lead to novel predictions that may soon be testable using the observational data of gravitational waves \cite{spinny,collision,QLove,CLove}.

One notable drawback of using the polymer model to describe ``real-world'' astrophysical BHs is that its strongly non-classical state implies that the interior  is lacking a semiclassical description \cite{BruVen,density,noclass}.
Moreover, the Einstein equations have no room for entropy, which is rather awkward when it comes to describing a gravitating matter system whose defining feature is its highly entropic status. Thus the polymer model cannot be  used  to deal directly with questions of a geometric nature.

In \cite{bookdill} (also see \cite{BHfollies}), we proposed a strategy for a classical description that captures important aspects
of the polymer model, but as it would be viewed by someone who is ignorant about the   importance
of quantum mechanics in describing gravitational collapse.
To elaborate on the strategy, let us first recall that the polymer model has the largest possible pressure $p$
that is allowed by causality $\;p=\rho\;$ ($\rho$ is the energy density), as
this condition translates into the desired feature of maximal entropy
$\;s=\frac{1}{T}(\rho+p)\;$ ($s$ is the entropy density and $T$ is the temperature).
 But an observer who knows nothing about the quantum nature of the interior
will  set the  entropy density $s$ to zero. And so  what
once was  a maximally positive pressure  has now become  maximally negative,
$\;p=-\rho\;$, when viewed from this classical perspective. Indeed,
maximally negative pressure is just what is needed to evade the singularity theorems \cite{PenHawk1,PenHawk2}, as well as the
Buchdahl bound \cite{Buchdahl} and similar limits
\cite{chand1,chand2,bondi}. See \cite{MM} for further discussion on this point.

The frozen star shares some similarities with other UCOs that exploit large negative pressure such as the black star \cite{barcelo}, the gravastar \cite{MMfirst} and even a (sort of) hybrid of the two \cite{CR}.  More specifically, the frozen star can be viewed as a limiting case of the black
star, whereas the main distinction between the  gravastar and our model has
to do with the (an)isotropy of the pressure. Unlike the isotropic nature of
the gravastar pressure, our model has a maximally negative radial component
$p_r$,
whereas the other (transverse) components $p_{\perp}$ are regarded as vanishing, at least on average. Having a  preferential spatial coordinate is quite natural given the spherical symmetry of the frozen star.

The geometry of the frozen star, although regular, is quite unusual
(however, see \cite{guendel1,guendel2}).
It has the especially peculiar feature that $\;g_{tt}=g^{rr}=0\;$ at all points
within the interior. This can again be traced to its association with
the polymer model for which maximal entropy means that the  Bekenstein--Hawking entropy bound \cite{Bek,Hawk} is saturated at every radius less than or equal to
the Schwarzschild radius.
In other words, every spherical shell from the center up to the outer surface
behaves just like
a horizon.

The  combination  of $\;g_{tt}=g^{rr}=0\;$ and $\;p_r+\rho=0\;$  is rather restrictive; in fact, these conspire to prescribe  a background solution that fails to support any form of small fluctuations of the metric and the matter densities, justifying the term frozen star.
This was evident from an inspection of the linearized Einstein equations, as elaborated on in \cite{bookdill}.

With a hat tip to nostalgia, we will  now refer to  this classicalized version
of the polymer BH as a ``frozen star''.

An important but previously unaddressed question is what exactly does  happen at the
outermost layer of the collapsed polymer/frozen star model. This would be difficult to answer from the polymer perspective, but the expectation is that the energy density and pressure decrease in a continuous way over a string-scale distance from their  interior values to zero, as does the entropy density. Translated to
the classical frozen star model, the geometry can be expected to interpolate
smoothly between the exotic interior and the Schwarzschild exterior.   This is, however,  far from being trivial as the smoothing has to  happen over a small-enough length scale to ensure the persistence of the model's  main features. But even if such a solution does exist, there is the additional question of whether or not perturbative stability is maintained
when this type of transitional layer is incorporated \cite{0310107}. There are also some concerns about the extremely large transverse pressures that are part and parcel with this type
of setup \cite{0505137}.~\footnote{Note, though, that large anisotropies
in the pressure act to weaken the Buchdahl bound, if anything \cite{9903067},
so that the relevant findings in \cite{bookdill}
are expected to remain valid.}  Our current objective is to address these issues
by using the classically geometric framework  of the frozen star model.

The  presentation proceeds as follows: First, we confirm the existence of
a suitable metric for the  transitional layer near the outer boundary.
This metric is obtained  by matching an appropriate ansatz to both
the bulk interior and the Schwarzschild exterior at appropriate
interfaces,
while insisting on
the continuity of various geometric quantities and matter fields.
Many of the details of this procedure
and some  supporting analysis  are  deferred to an appendix. Next,
the important issue of stability is addressed. We conclude with a discussion
on why large transverse pressures in the transitional layer are no cause for
concern, followed by a brief overview of our results.

\subsection*{Conventions}

We assume a spherically symmetric and static background spacetime
with  $\;D=3+1\;$ spacetime  dimensions, although a similar analysis
and  conclusions will  persist for any $\;D>3\;$. All  fundamental constants besides Newton's constant $G$ are set to unity throughout. In the stability analysis, we further fix
$\;8\pi G=1\;$. A prime (dot) indicates a radial (temporal) derivative.

\section{A metric for the transitional layer}

\subsection{The bulk}

Before discussing the crust --- the thin transitional layer between the bulk and the boundary --- we first recall some basics about the interior bulk of the frozen star. Let us begin with a static and spherically symmetric
line element,
\be
ds^2\; =\; -f(r) dt^2 + \frac{1}{{\widetilde f}(r)} dr^2 + r^2 (d\theta^2+ sin^2 \theta d\phi^2)\;.
\label{linement}
\ee
It is then  assumed (as discussed above) that the radial pressure is maximally negative,  $\;p_r=-\rho\;$. The transverse components  $p_{\perp}$ are, on the other hand, left unspecified for the time being. All of the off-diagonal elements
of the stress tensor are vanishing.

Under these conditions,  Einstein's equations reduce to
\begin{align}
\label{E1}
\left(r{\widetilde f}\right)'  & \;=\; 1-8\pi G\rho r^2\;, \\
\label{E2}
\left(rf\right)''\; & \;=\; 16\pi G r p_{\perp}\;.
\end{align}
Given that $\;f={\widetilde f}\;$, the previous pair are equivalent to
\be
\label{E3}
\left(\rho r^2\right)'  \;=\; - 2 r p_{\perp}\;,
\ee
which is then consistent with the  conservation
of the stress tensor.

Let us next define
\be
\label{mofr}
m(r)\;=\;4\pi\int\limits_0^r dx\, x^2 \rho(x)\;\;\; {\rm for} \;\;\; r\leq R\;.
\ee
One then finds that
\be
f(r)\;=\; {\widetilde f}(r) \;=\; 1- \frac{2 m(r)}{r}\;.
\ee

The functional form
of $\rho(r)$ or, equivalently, $m(r)$, is what  ultimately determines
the geometry of the UCO. For instance, the gravastar solution is determined by the
choice  $\rho={\rm const}.$; in other words, a de Sitter interior. As the frozen star should be the classical analogue of the polymer model --- for which the entropy--area law is saturated throughout ---
the natural choice is to saturate the Schwarzschild limit $\;m(r)=r/2\;$, also throughout.
Equivalently, $\;\rho\sim 1/r^2$. We thus end up with the advertised outcome
of $\;f=0\;$ and the matter densities can be shown to adopt the following profiles:
\bea
\label{polyrho}
8\pi G  \rho &=& \frac{1-(rf)'}{r^2} \;=\;  \frac{1}{r^2}\;, \\
\label{polypr}
8\pi G  p_r &=& -\frac{1-(rf)'}{r^2} \;=\; -\frac{1}{r^2}\;, \\
\label{polypt}
8\pi G  p_\perp &=& \frac{(rf)''}{2r}\;=\; 0\;.
\eea

\subsection{The crust}

We now want to implement the concept of a transitional layer in concrete terms.
The UCO is regarded as having a radius of $R$, $\;r=R\;$  being the midpoint of
a thin transitional layer. This layer will be taken to range from $\;r=r_A=R-\lambda\;$ to $\;r=r_B=R+\lambda\;$, where $\;\lambda \ll R\;$. Because of the star's connection to the polymer model, we expect that $\lambda$ is of order of the string length $l_s$,  $\;\lambda \sim
l_s\;$, but its exact value is not important to the current work. What is important is that $\lambda$ is parametrically larger than the Planck length, so that the Einstein equations remain valid  in the crust.

We start with the assumption that the key symmetries of the frozen star interior,
$\;{\widetilde f}=f\;$ and $\;p_r=-\rho\;$,  persist into the transitional layer. The form of the metric  is found by adopting the ansatz that $f(r)$ is a polynomial expansion in terms of $\;\frac{r-r_A}{R}<\frac{2\lambda}{R}\;$. The order of the polynomial and, thus,
the number of  adjustable parameters in the ansatz  is determined by the number of relevant boundary conditions. For this analysis, we are imposing that $f(r)$, $f'(r)$ and $f''(r)$ be continuous at both ends of the layer. This means the imposition of the matching conditions $\;f=f'=f''=0\;$ at $\;r=r_A\;$, as well as  matching $f$ and its first two derivatives to their standard Schwarzschild values at $\;r=r_B\;$. This procedure results in a fifth-order polynomial, as described in the Appendix. Additional conditions could be imposed, if one so desires, but at the cost of additional parameters. We were satisfied with these six because they were enough to ensure the continuity of all the stress-tensor components, as well as their first derivatives,  at both ends of the layer.
In Fig.~{\ref{Fig-f}, we depict the metric function and the stress-tensor components in the transitional layer for various values of $\lambda$.

\begin{figure}[t]
\vspace{-.2in}
	\includegraphics[scale=0.3]{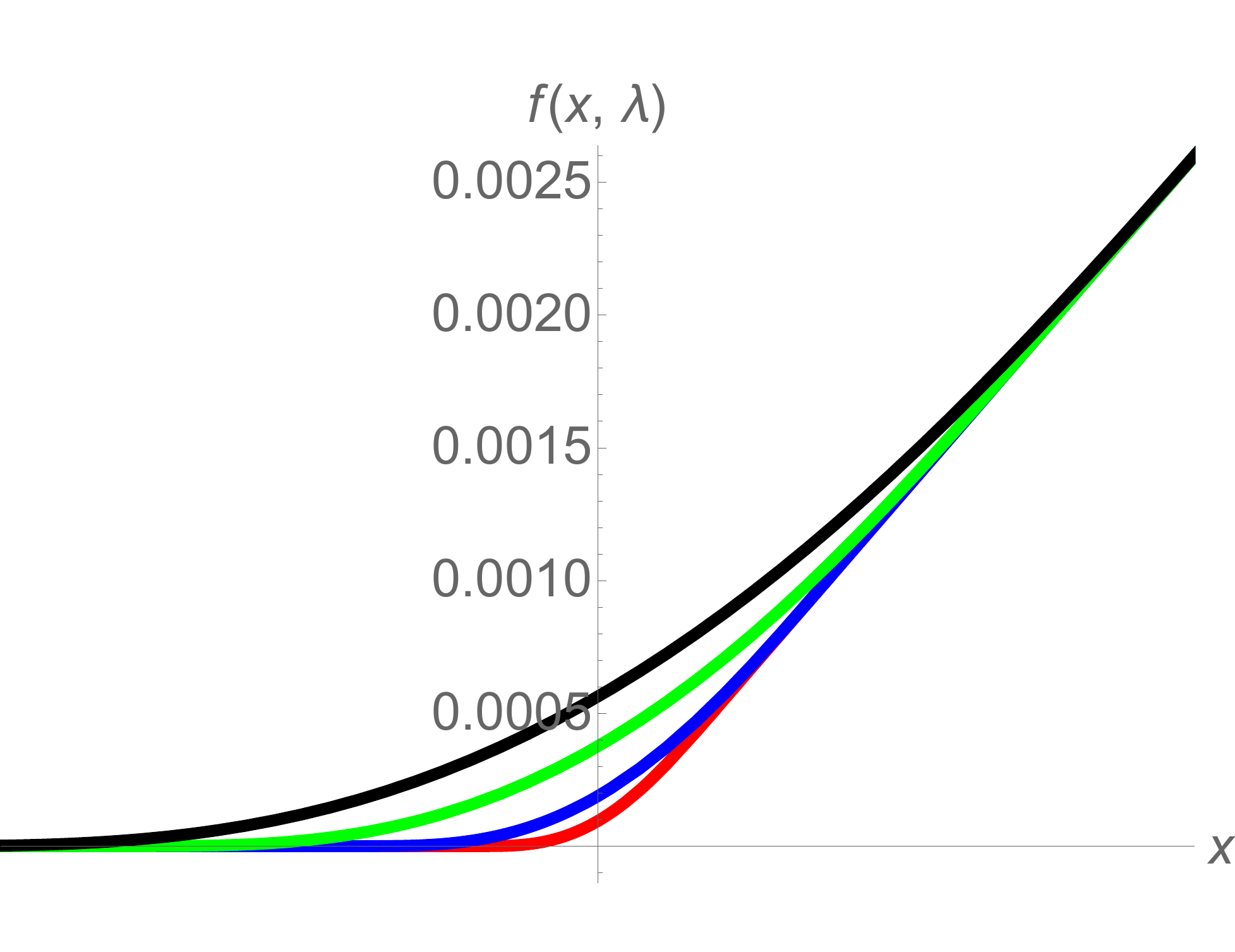}\hspace{0.5in}\includegraphics[scale=0.3]{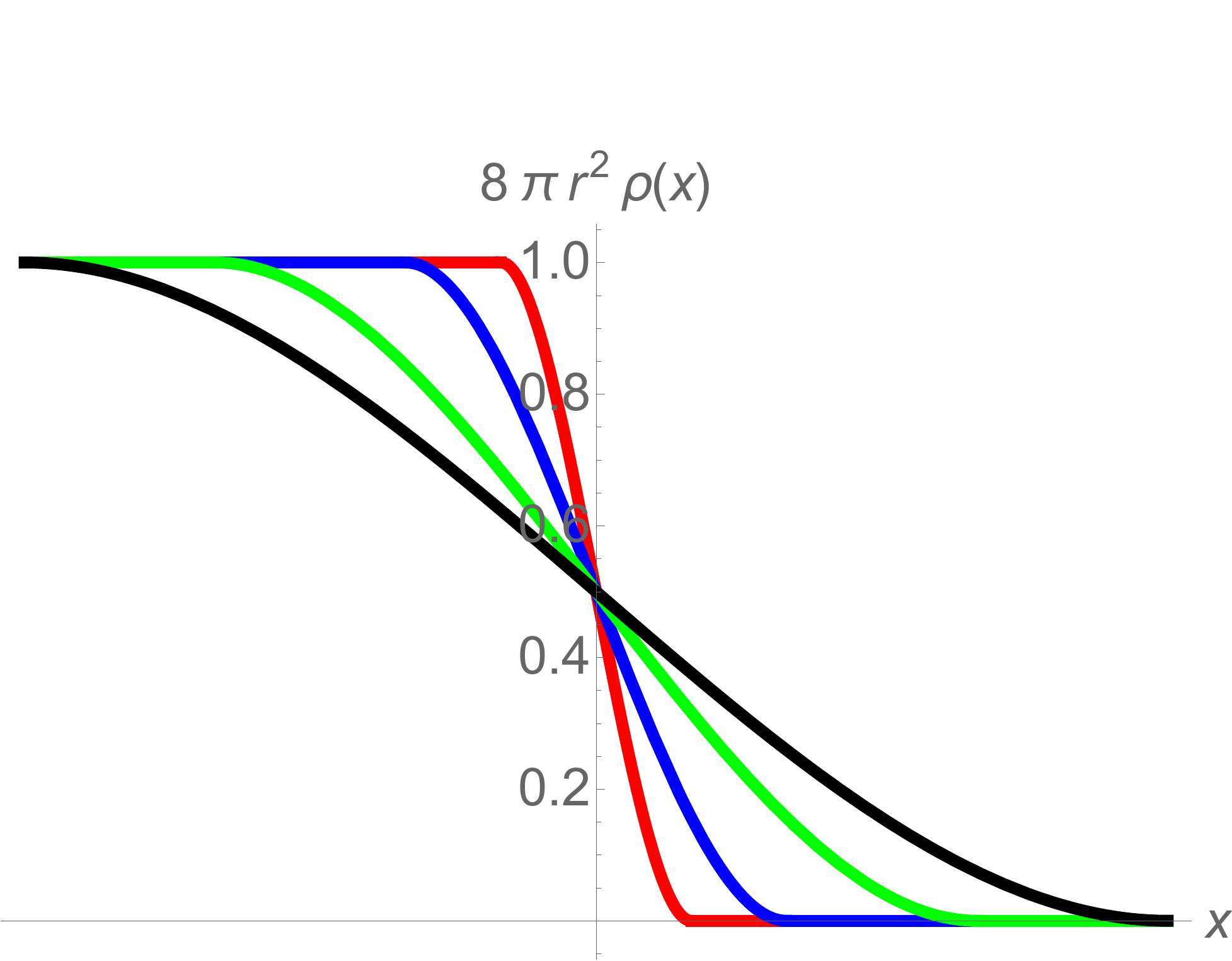}\\
\includegraphics[scale=0.3]{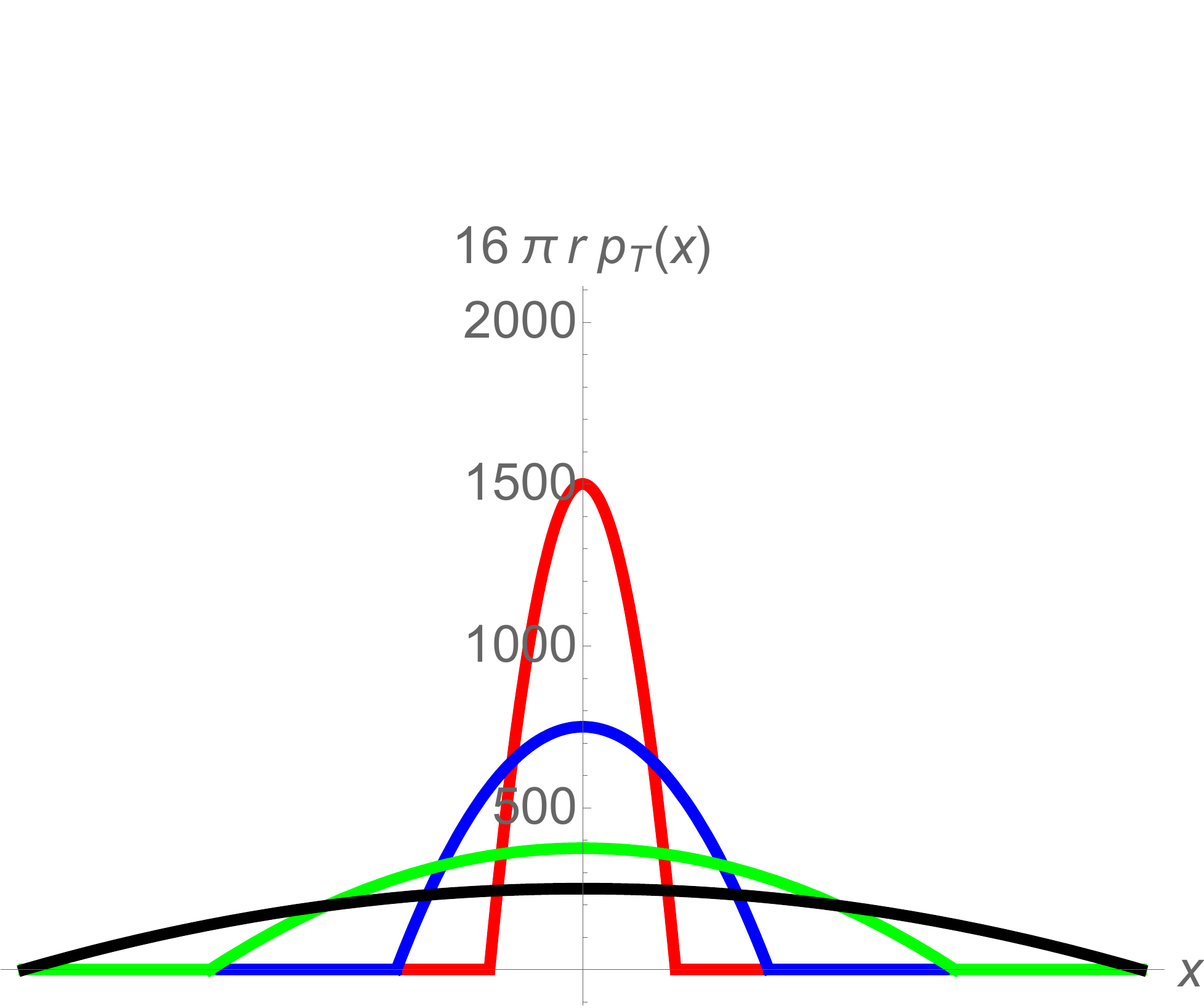}\hspace{0.5in}\includegraphics[scale=0.3]{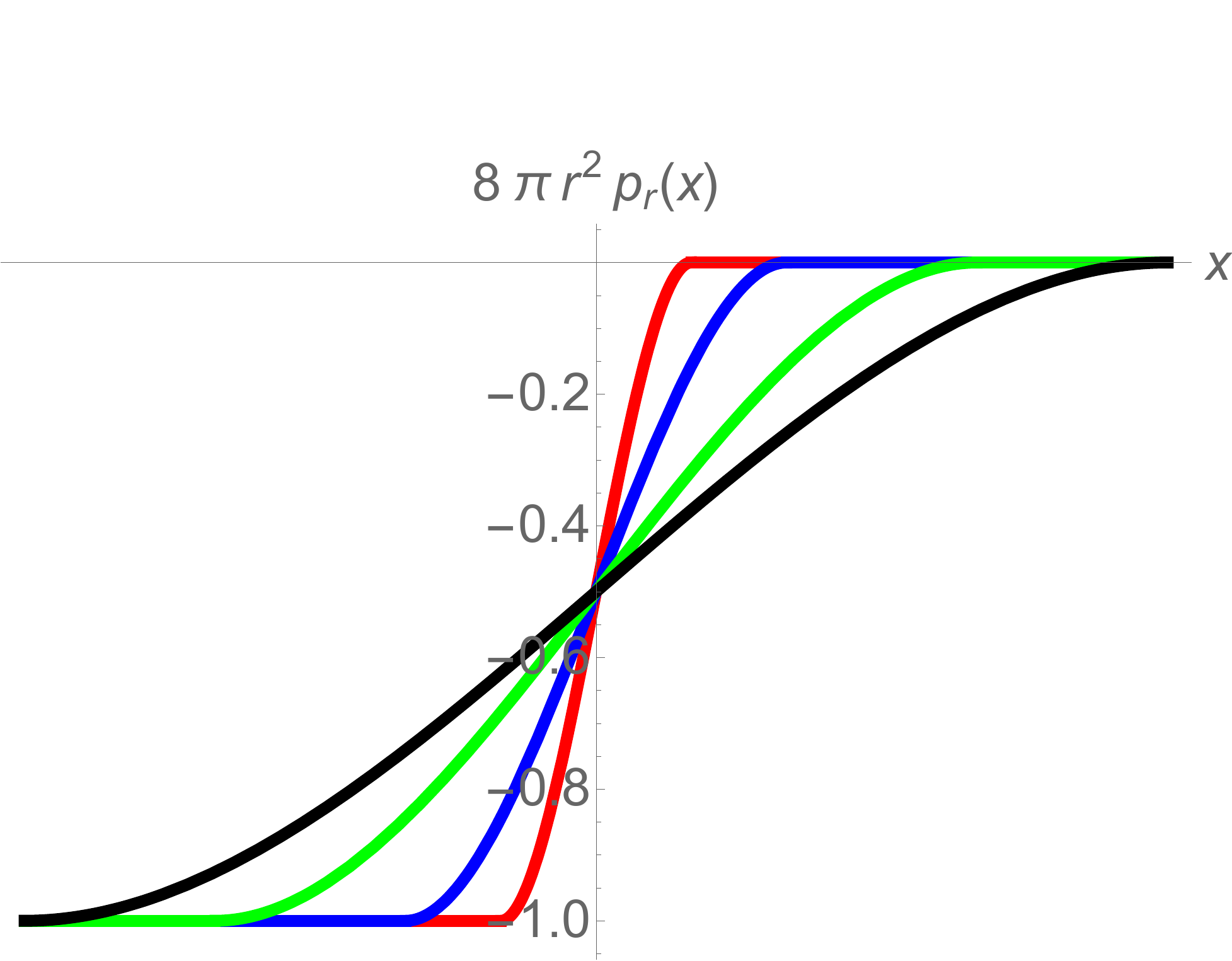}
	\caption{The panel depicts, for several values of the dimensionless width parameter $\lambda/R$, the metric function $f$ (upper left), the energy density (upper right), the transverse pressure (lower left) and the radial pressure (lower right) as a function of the dimensionless parameter $x=r/R$. The vertical axis is positioned at $x = 1$ in the middle of the transitional layer, $ x < 1 $ corresponds to the part of the transitional layer closer to the bulk and $x>1$ to the part closer to the exterior.}
	\label{Fig-f}
\end{figure}

In spite of the simplicity of the methodology, the actual expressions  are quite complicated and so have been relegated
to the Appendix along with some supporting analysis.  What is worth emphasizing is that $f$ and $\rho$ are non-negative throughout the spacetime ({\em cf},
Sections~A.6 and~A.7). The same
can be said about $p_{\perp}$ (see A.8),
ensuring that the null energy condition is never violated.
It is also worth a mention that, in the layer itself, $\;\frac{1}{r}(r^2 \rho)'\sim
-\frac{R}{\lambda}\rho\;$, meaning that
$\;p_{\perp} \sim \frac{R}{\lambda}\rho
\gg \rho\;$;  {\em cf}, Eq.~(\ref{E3}).~\footnote{This last relation and
the mutual $1/\lambda$ scaling is confirmed directly in Section~A.8.}
A large transverse pressure is an inevitable consequence of
trying to increase  a maximally  negative $p_r$ to zero over a small length scale,
as first elaborated on within \cite{0505137} in reference to the gravastar model.
Why this is not really a bad thing for our model  will be the focus of the fourth section.

\section{Linear ultra-stability}

It was shown in  \cite{bookdill} that, for the frozen star, all linear fluctuations about the background solution are vanishing; the star is not just  stable
but   ``ultra-stable''. This  outcome was not unexpected because of an analysis
in \cite{emerge} (following a similar one in \cite{LT}) which  revealed that perturbations of the polymer's equilibrium state die out exponentially quickly. But, in the case of
the frozen star, the stability can best be attributed to  a pair of strong
constraints on the background solution; namely,
$\;p_r+\rho=0\;$ and $\;f=-g_{tt}=g^{rr}=0\;$. However, the  status of the latter changes in the transitional layer as
the metric function $f(r)$ is no longer vanishing. Still, it will  be
shown  that the  ultra-stability persists by virtue of
the continuity of the metric
through the interface at $r=r_A$.
To this end, we will be following
Chandrasekhar's linear stability analysis \cite{chand1,chand2}.

For the rest of this section, the zeroth-order matter functions carry a subscript of 0  and $\;8\pi G=1\;$.

The metric and stress tensor are now expressible as
\bea
g_{\mu\nu} &\;=\;& \text{diag}
\Biggl( -f(r)+\delta g_{tt}(t,r),\ \ 1/f(r) +\delta g_{rr}(t,r),\ \  r^2,\ \  r^2 \sin^2\theta\Biggr)\; \cr
&\;=\;&\text{diag}
\Biggl( -f(r) (1-H_0(t,r)),\ \ 1/f(r) (1+ H_2 (t,r)),\ \  r^2,\ \  r^2 \sin^2\theta\Biggr)\; \ \ \ \ \
\label{metricX}
\eea
and
\be
T^\mu_{\;\;\nu}\;=\; \text{diag}
\Biggl( -{\rho}_0-\delta\rho, \ \ -{\rho}_0+\delta p_r,\ \  {p}_{\perp}{}_{0}+\delta p_{\perp},\ \ {p}_{\perp}{}_{0}+\delta p_{\perp}\Biggr)\;.
\ee

Let us next consider, one by one, the perturbed forms of the (non-trivial) Einstein equations, as well as
the perturbed conservation equation. We are assumed to be working
strictly  in the transitional layer
as defined in the previous section, $\;R-\lambda \leq r \leq R+\lambda\;$ or
$\;r_A \leq r \leq r_B\;$.

The ${}^t_{\ r}$ equation is
\be
\dot{H}_2(t,r)\;=\; \tfrac{r}{f(r)}\left(\rho_0+p_r{}_0\right) \text{v}_r\;,
\ee
where $\;\text{v}_r=\dot{r}\;$ is the radial velocity and we use this to
further define
the Lagrangian displacement $\xi$ such that  $\;\text{v}_r= \partial_t \xi\;$.
Integrating this equation with respect to time, one then obtains
\be
H_2(t,r)\;=\; \tfrac{r}{f(r)}\left(\rho_0+p_r{}_0\right) \xi\;.
\ee
But, since $\;\rho_0+p_r{}_0=0\;$, it immediately follows that
\be
H_2(t,r)\;=\;0\;.
\label{H2vanish}
\ee
This is consistent with the bulk analysis  \cite{bookdill} and continuity.
The conclusion is that $g_{rr}$ is frozen at its background value throughout the transitional crust, just as it is for the bulk interior.

The ${}^t_{\ t}$ equation takes the form
\be
-\frac{1}{r^2}+ \frac{f+rf'}{r^2}- \frac{f+rf'}{r^2} H_2(t,r)-\frac{f}{r^2}H'_2
\;=\;-\rho_0-\delta\rho(t,r)\;.
\ee
Identifying $\rho_0$ from Eq.~(\ref{polyrho})  and using the vanishing of $H_2$, we find that $\delta\rho$ is also frozen,
\be
\delta\rho(t,r)\;=\;0\;.
\label{drhovanish}
\ee

The ${}^r_{\ r}$ equation goes as
\be
-\frac{1}{r^2}+ \frac{f+rf'}{r^2}- \frac{f+rf'}{r^2} H_2(t,r)-\frac{f}{r}H_0'
\;=\;p_r{}_0+\delta p_r(t,r)\;.
\ee
Identifying $p_r{}_0$ via Eq.~(\ref{polypr})  and taking  $H_2$ to zero,
we have
\be
- \frac{f}{r}H_0'\;=\;\delta p_r\;.
\label{deltapr}
\ee

Meanwhile, the ${}^\theta_{\ \theta}$ equation leads to
\bea
-\frac{1}{2 f}\partial_t^2 H_2&+& \frac{2f'+rf''}{2 r}- \frac{2f'+rf''}{2 r} H_2(t,r)-\frac{f'}{4}(3H_0'+H_2')\cr &-&\frac{1}{2 r}(H_0'+H_2'+r H_0'')\;=\; p_{\perp}{}_0
+\delta p_{\perp}(t,r)\;.
\eea
With the identifications of  $p_{\perp}{}_0$ from Eq.~(\ref{polypt})
and $\;H_2=0\;$,
\be
-\frac{f}{2r}H_0'-\tfrac{3}{4} f' H_0' -\frac{f}{2}H_0''\;=\;\delta p_{\perp}(t,r).
\label{deltapT}
\ee

As for the conservation equation,  once  $\;H_2=0\;$, $\;\delta\rho=0\;$
and the zeroth-order terms are all accounted for, this becomes
\be
\nabla_\mu T^\mu_{\ r}\;=\; -\frac{2}{r}(\delta p_r-\delta p_{\perp})+\tfrac{1}{2}\frac{f'}{f} \delta p_r+ \delta p_r'=0\;.
\label{conservative}
\ee
Equation~(\ref{conservative}) can be turned, using Eqs.~(\ref{deltapr})
and~(\ref{deltapT}), into a differential equation for $H_0$ alone,
\be
2\frac{f}{r^2}H_0'- 3 \frac{f'}{r} H_0'-2\frac{f}{r}H_0''\;=\;0\;.
\label{h0eq}
\ee

The most general solution for  Eq.~(\ref{h0eq}) is readily obtained,
\be
H_0'(t,r)\;=\; A(t) e^{\ \int\limits_{r_A}^{r}\left(\frac{3}{2}\frac{f'}{f}-\frac{1}{r}\right) dr} +B(t)\;,
\ee
where  $A(t)$ and $B(t)$ are integration ``constants'' which can be fixed
at the inner boundary of the transitional region, $\;r=r_A\;$.

Recall from the prior section that the  boundary conditions at $r_A$ are
$\;f=f'=f''=0\;$. Technically, these conditions apply only to the background
geometry. But on this we can say  more because $r_A$ is also part of the bulk
interior where the fluctuations are all vanishing. So that, by continuity,
$\;H_0=H_0'=H_0''=0\;$ are  also true at the  $\;r=r_A\;$ interface.
Let us first impose  $\;H_0'(r_A)=0\;$ to obtain $\;B=-A\;$ and then  $\;H_0''(r_A)=0\;$ which leads to  $\;A=0\;$, meaning that  $\;H_0'=0\;$
for all $r$.  The result is that $H_0$ must be equal to a constant
and, by further imposing $\;H_0(r_A)=0\;$,
we know that the constant in question is
zero. This leads to the  conclusion that  $H_0$, just like $H_2$,
is frozen throughout,
\be
H_0\; =\;0\;.
\label{h0vanish}
\ee

Finally,   Eqs.~(\ref{deltapr}) and~(\ref{deltapT}) now tell us that  both
types of  pressure perturbations are vanishing, just like $\delta \rho$
was shown to be.

To summarize,  we have found that  all metric and matter perturbations
vanish identically even when the frozen star has been  topped off
with an outer layer for which $\;f(r)\neq 0\;$.

\section{Causality}

As first discussed  in  \cite{0505137} and noted above, any UCO model  that is premised on the idea of a radial pressure that is large and negative will inevitably lead to very large transverse pressures in the boundary layer. An easy way to see this is to reconsider the stress-tensor conservation equation but with our previous condition of  $\;\rho+p_r=0\;$ now being  relaxed,
\be
p'_r + \frac{f'}{f}(\rho+p_r)+\frac{2}{r}(p_r-p_{\perp})=\;0\;\;.
\label{conservy}
\ee
For a  negative radial pressure whose magnitude  is the same order as the energy density, the middle term on the left-hand-side of Eq.~(\ref{conservy}) is parametrically smaller than the other two terms. Meanwhile, this large negative radial pressure has to increase to zero at the outer edge of a relatively
thin transitional layer, thus making its radial derivative large. The transverse pressure is then left to compensate for this large value of $p'_r$. The good news is that the transverse pressure must be positive to do its job, so that the null energy condition is never violated. There is, however, the  issue of causality, as well as the physical interpretation of large transverse pressures  for a static configuration with no external forces.

Let us start with the apparent violation of causality because of
$\;p_{\perp} \sim \frac{R}{\lambda}\rho \gg \rho\;$ in the transitional layer. Fortunately, there are two ways --- one local and one non-local --- to see that $p_{\perp}\gg\rho$ does not translate, as one might perhaps expect, into sound speeds in excess of the speed of light. And so,  in spite of appearances. the  crust does not support superluminal propagation.

The local way is to calculate the transverse speed of sound $c_S{}_{\perp}$  via the
thermodynamic relation
$\;c^2_S{}_{\perp}=\left|\frac{\delta p_{\perp}}{\delta\rho}\right|\;$. This can be accomplished by
applying the Euler---Lagrange equation directly to Eq.~(\ref{conservy}),
along with  the frozen star  equation of state, $\;p_r=-\rho\;$,
\be
-\partial_r \frac{\delta p_{r}}{\delta\rho}+ \frac{f'}{f}(1+\frac{\delta p_{r}}{\delta\rho})+\frac{2}{r}\left( \frac{\delta p_{r}}{\delta\rho} - \frac{\delta p_{\perp}}{\delta\rho}\right)
\;=\;0\;.
\ee
The variation $\;\frac{\delta p_{r}}{\delta\rho}=-1\;$ leads to $\;c^2_S{}_{\perp}=1\;$, which is what a hypothetical local observer would measure for the speed of transverse-moving modes.

The non-local way of calculating the transverse speed of sound is to use the line element for the transitional  layer,
$\; c^2_S{}_{\perp}=r^2\frac{d\theta^2}{dt^2}=f(r)\;$.  In this layer, a typical value of $f(r)$  is small but nonvanishing, $\;f(r) \sim \frac{\lambda}{R}  \;$, which
follows from $f(r)$ varying from 0 up to $\lambda/(R+\lambda)$  as per  Eqs.~(\ref{second conR}) and~(\ref{first conR}) in the Appendix. This
result  for the transverse speed is what a distant observer would measure. However, to compare with the previous result, one must take the effect of the gravitational redshift into account,
which essentially  means dividing the transverse speed of sound  $c_S{}_{\perp}$  by $\sqrt{f(r)}$, leading to $\;c^2_S{}_{\perp}=1\;$. Hence, the two calculations are in complete agreement, and we can  conclude that there is no threat to causality.

It is tempting to  suggest that the large transverse pressures are merely a mathematical artifact of maintaining energy conservation while inside the transitional layer. To support this claim, we can turn
to the polymer perspective, where a maximally negative
value of $p_r$ is, itself, a fictitious consequence of ignoring the maximally entropic state  of the internal matter. Indeed, from this stringy point of view,
one can set $\;p_{\perp}\approx 0\;$ throughout the transitional layer and can replace $p_{r}+\rho$ with $sT$ (again, approximately).
The conservation equation~(\ref{conservy}) now looks like
\be
p'_r + \frac{f'}{f}(sT)+\frac{2}{r}p_r\;=\;0\;.
\label{conservy2}
\ee
In the polymer model, it is a maximally positive $p_r$ that must decline rapidly to zero
in moving outwards through the transitional layer, so that $p'_r$ is
negative and  of order
$\rho/\lambda$. But this is exactly the same order as  the (manifestly positive) entropic term. There is no need for a transverse pressure, large or otherwise,  when viewed from this perspective.

\subsection*{Overview}

We have shown that our   frozen star model --- which can be regarded as the geometrical manifestation of  the  polymer model for the BH interior ---
can readily incorporate a thin transitional layer at the outer surface of a
so-described UCO. Notably, this outer crust  maintains the same ultra-stability
of the interior bulk and, furthermore, poses no threat to  causality or
the null energy condition.

These  results could further our understanding on just how
the collapsed polymer/frozen star model modifies the experimental signatures of
gravitational waves emitted during a binary BH merger and of out-of-equilibrium physics from those of  the general relativistic paradigm. See \cite{ridethewave,spinny,collision,QLove,CLove} for  progress along these lines from a somewhat different perspective.

\section*{Acknowledgments}

The research of AJMM received support from an NRF Evaluation and Rating Grant
119411.
The research of RB and TS was supported by the Israel Science Foundation grant no. 1294/16.

\begin{appendix}

\section{The transitional layer in detail}

The purpose of this Appendix is to give a detailed account of
how we found the metric for the transitional layer. Also included
is analysis in support of various statements that are made
in the main text of the paper.

\subsection{The model}

We start here by recalling that our model  --- the frozen star --- describes a
4-dimensional, static, spherically symmetric UCO of radius $R$, for which the radial component of the pressure
is maximally negative, $\;p_r=-\rho\;$. Let us also revisit the line element~(\ref{linement}), but now with the knowledge that $\;-g_{tt}=g^{rr}\;$ is required for consistency with the  conservation of the  stress tensor,
\begin{equation}
	ds^2\;=\;-f(r)dt^2+\frac{1}{f(r)}dr^2+r^2(\sin^2\theta+d\phi^2)\;.
\label{metric}
\end{equation}

Also recalled, from Section~2, are
the expressions for the energy density, radial pressure and transverse pressure,
\begin{equation}
	\rho(r)\;=\;\frac{1-f-rf'}{8\pi G r^2}\;,
\end{equation}
\begin{equation}
	p_r(r)\;=\;-\frac{1-f-rf'}{8\pi G r^2}\;,
\end{equation}
\begin{equation}
	p_\perp(r)\;=\;\frac{2f'+rf''}{16\pi Gr}\;.
\end{equation}

Finally, let us reiterate  our proposed framework:  The sphere is assumed
to have an outermost transitional layer of thickness $\;2\lambda\ll R\;$,
centered around $\;r=R\;$.
That is, the layer is half inside and half outside of the sphere, as is conventional in problems
involving the transition between two media.
The  inner boundary of the layer at $\;r=R-\lambda\;$ connects to the bulk interior
(for which
$f$ and all its derivatives vanish  throughout) and its outer boundary
at $\;r=R+\lambda\;$ connects to the Schwarzschild
exterior.  Effectively invoking Occam's razor, we assume that the interior  conditions $\;p_r+\rho=0\;$
and $\;g_{tt}+g^{rr}=0\;$  persist into the layer.

\subsection{Continuity conditions for \texorpdfstring{$f(r),\; f'(r),\; f''(r)$}{Lg}}

Let us begin  here by matching $f$ and its first two derivatives at each of
the two endpoints of the  transitional layer. It can be checked that
the continuity  of these three functions is sufficient to ensure the continuity  of the matter
densities as well.
To this end, we now introduce a function $S(r,\lambda)$ (with the $\lambda$
sometimes implied) that will be used to express $f$ in the layer,
\begin{equation}
	f\;=\;\begin{cases}
		0,\quad r<R-\lambda\\
		S(r), \quad R-\lambda<r<R+\lambda\\
		1-\frac{R}{r}, \quad r>R+\lambda\;,
	\end{cases}
\end{equation}

\begin{equation}
	f'\;=\;\begin{cases}
		0, \quad r<R-\lambda\\
		S'(r), \quad R-\lambda<r<R+\lambda\\
		\frac{R}{r^2}, \quad r>R+\lambda\;,
	\end{cases}
\end{equation}

\begin{equation}
	f''\;=\;\begin{cases}
		0, \quad r<R-\lambda\\
		S''(r), \quad R-\lambda<r<R+\lambda\\
		-\frac{2R}{r^3}, \quad r>R+\lambda\;.
	\end{cases}
\end{equation}

The matching conditions  on $S(r,\lambda)$ at the layer's endpoints
are then as follows:

For the continuity of $f$,
\begin{equation}
	1. \quad S(r=R-\lambda,\lambda)\;=\;0\label{second conR}\;,
\end{equation}
\begin{equation}
	2.\quad S(r=R+\lambda,\lambda)\;=\;\frac{\lambda}{R+\lambda}\label{first conR}\;.
\end{equation}

For the continuity of $f'$,
\begin{equation}
	3. \quad S'(r=R-\lambda,\lambda)\;=\;0\;,\label{fourth conR}
\end{equation}
\begin{equation}
	4. \quad S'(r=R+\lambda,\lambda)\;=\;\frac{R}{(R+\lambda)^2}\;.\label{thirsCon}
\end{equation}

And, for the  continuity of $f''$,
\begin{equation}
	5. \quad S''(r=R-\lambda,\lambda)\;=\;0\label{last conR}\;,
\end{equation}
\begin{equation}
	6. \quad S''(r=R+\lambda,\lambda)\;=\;-\frac{2R}{(R+\lambda)^3}\label{fifth conR}\;.
\end{equation}

It is convenient at this point to introduce a
dimensionless radial coordinate, $\;x=\frac{r}{R}\;$,
so that the transitional layer $\;R-\lambda<r<R+\lambda\;$
can be alternatively defined as $\;1-\lambda<x<1+\lambda\;$.~\footnote{In terms of the new coordinate,
$\lambda$ really means $\;\tilde{\lambda}=\lambda/R\;$, but we will not bother
to make this distinction.} In what follows, a prime indicates
a radial derivative with respect to  its argument and a derivative with respect
to $r$ when no argument is provided.

\subsection{Continuity conditions for \texorpdfstring{$\rho(r),\; p_r(r),\; p_\perp(r)$}{Lg}}

The energy density, radial pressure, and transverse pressure are
re-expressed below in terms of the function $S(r,\lambda)$.
One can readily  verify our claim  that
the continuity of  $\rho(r),\; p_r(r),\;{\rm and}\; p_\perp(r)$
through the endpoints of the translational layer
is consistent  with the conditions~(\ref{second conR})-(\ref{fifth conR}).

\begin{equation}
	\rho\;=\;\begin{cases}
		\frac{1}{8\pi G r^2}, \quad r<R-\lambda\\
		\frac{1-S(r)-rS'(r)}{8\pi G r^2}, \quad R-\lambda<r<R+\lambda\\
		0, \quad r>R+\lambda\;,
	\end{cases}\label{rho}
\end{equation}

\begin{equation}
	p_r\;=\;\begin{cases}
	-	\frac{1}{8\pi G r^2}, \quad r<R-\lambda\\
		-\frac{1-S(r)-rS'(r)}{8\pi G r^2}, \quad R-\lambda<r<R+\lambda\\
		0, \quad r>R+\lambda\;,
	\end{cases}\label{pr}
\end{equation}

\begin{equation}
	p_\perp\;=\;\begin{cases}
		0, \quad r<R-\lambda\\
		\frac{2S'(r)+rS''(r)}{16\pi G r}, \quad R-\lambda<r<R+\lambda\\
		0, \quad r>R+\lambda\;.
	\end{cases}\label{pt}
\end{equation}


\subsection{Continuity conditions for \texorpdfstring{$\rho'(r),\; p'_\perp(r)$}{Lg}}

The expressions for the first derivatives  of the  energy density
and the  transverse pressure take on  the respective forms,~\footnote{We have omitted $p'_r(r,\lambda)$ given that it is identically the negative of $\rho'(r,\lambda)$.}

\begin{equation}
	\rho'(r,\lambda)\;=\;\begin{cases}
		-\frac{2}{8\pi G r^3}, \quad r<R-\lambda\\
		-\frac{2-2S(r)+r^2S''(r)}{8\pi G r^3}, \quad R-\lambda<r<R+\lambda\\
		0, \quad r>R+\lambda\;,
	\end{cases}
\end{equation}

\begin{equation}
	p'_\perp(r,\lambda)\;=\;\begin{cases}
		0, \quad r<R-\lambda\\
		\frac{1}{16\pi G}\left(-\frac{2S'(r)}{r^2}+\frac{2S''(r)}{r}+S'''(r)\right), \quad R-\lambda<r<R+\lambda\\
		0, \quad r>R+\lambda\;.
	\end{cases}
\end{equation}

In order for $\rho'(r,\lambda)$ to be continuous, what is required is that

1.\quad$  2-2S(R-\lambda)+(R-\lambda)^2S''(R-\lambda)\;=\;2$\;,

2.\quad$  2-2S(R+\lambda)+(R+\lambda)^2S''(R+\lambda)\;=\;0$\;.

The former is satisfied as a consequence of  conditions~(\ref{second conR}) and~(\ref{last conR}), while the latter can be confirmed  using
conditions~(\ref{first conR}) and~(\ref{fifth conR}).

As for the first derivative of the transverse pressure, the matching conditions are  $\;S'''(r=R-\lambda,\lambda)=0\;$ and
$\;S'''(r=R+\lambda,\lambda)=\frac{6\lambda}{(R+\lambda)^4}\;$, where
the conditions~(\ref{fourth conR})-(\ref{fifth conR}) have been employed.
Insisting on these conditions, one would need to start
with a polynomial of degree seven. We choose not to do so in the current analysis.


\subsection{Results as a function of \texorpdfstring{$x$}{Lg}}

Here, we report the main results as functions of the dimensionless radial coordinate $x$ , which can be
easily translated into functions of  $r$.

The function $f(x,\lambda)$ and its derivatives go as follows:
\bea
	f(x,\lambda)&=&\left(\frac{1+4\lambda+5\lambda^2}{4\lambda^2(1+\lambda)^3}\right)(x-1+\lambda)^3-\left( \frac{(1+3\lambda)(1+5\lambda)}{16\lambda^3(1+\lambda)^3}  \right)(x-1+\lambda)^4  \nonumber \\
&+&\left( \frac{1+3\lambda}{16\lambda^3(1+\lambda)^3}  \right)(x-1+\lambda)^5\;,
\eea
\bea
	f'(x,\lambda)&=&3\left(\frac{1+4\lambda+5\lambda^2}{4\lambda^2(1+\lambda)^3}\right)(x-1+\lambda)^2-4\left( \frac{(1+3\lambda)(1+5\lambda)}{16\lambda^3(1+\lambda)^3}  \right)(x-1+\lambda)^3  \nonumber \\
&+&5\left( \frac{1+3\lambda}{16\lambda^3(1+\lambda)^3}  \right)(x-1+\lambda)^4\;,
\eea
\bea\label{fx''}
	f''(x,\lambda)&=&6\left(\frac{1+4\lambda+5\lambda^2}{4\lambda^2(1+\lambda)^3}\right)(x-1+\lambda)-12\left( \frac{(1+3\lambda)(1+5\lambda)}{16\lambda^3(1+\lambda)^3}  \right)(x-1+\lambda)^2  \nonumber \\
&+&20\left( \frac{1+3\lambda}{16\lambda^3(1+\lambda)^3}  \right)(x-1+\lambda)^3\;.
\eea

Meanwhile, the energy density, radial pressure and  transverse pressure adopt
the following forms:

	\begin{eqnarray}
		(8\pi Gr^2)\rho(x,\lambda)&=&-\tfrac{(3 \lambda +1) (\lambda +x-1)^5}{16 \lambda ^3 (\lambda +1)^3}+\left(\tfrac{(3 \lambda +1) (5 \lambda +1)}{16
			\lambda ^3 (\lambda +1)^3}-\tfrac{5 (3 \lambda +1) x}{16 \lambda ^3 (\lambda +1)^3}\right) (\lambda+x-1)^4 \cr
		&+&\left(\tfrac{(3 \lambda +1) (5 \lambda +1) x}{4 \lambda ^3 (\lambda +1)^3}
-\tfrac{5 \lambda ^2+4 \lambda +1}{4 \lambda ^2 (\lambda +1)^3}\right) (\lambda +x-1)^3  \label{yikes} \\
&-&\tfrac{3 \left(5 \lambda ^2+4 \lambda +1\right) x (\lambda
			+x-1)^2}{4 \lambda ^2 (\lambda +1)^3}+1\;,  \nonumber
	\end{eqnarray}	

\begin{eqnarray}
			(8\pi Gr^2)p_r(x,\lambda)&=&\tfrac{(3 \lambda +1) (\lambda +x-1)^5}{16 \lambda ^3 (\lambda +1)^3}-\left(\tfrac{(3 \lambda +1) (5 \lambda +1)}{16
			\lambda ^3 (\lambda +1)^3}+\tfrac{5 (3 \lambda +1) x}{16 \lambda ^3 (\lambda +1)^3}\right) (\lambda+x-1)^4 \cr
		&-&\left(\tfrac{(3 \lambda +1) (5 \lambda +1) x}{4 \lambda ^3 (\lambda +1)^3}+\tfrac{5 \lambda ^2+4 \lambda +1}{4
			\lambda ^2 (\lambda +1)^3}\right) (\lambda +x-1)^3 \\
&+&\tfrac{3 \left(5 \lambda ^2-4 \lambda +1\right) x (\lambda+x-1)^2}{4 \lambda ^2 (\lambda +1)^3}-1\;,
\nonumber	
\end{eqnarray}

\begin{eqnarray}
\label{theperp}
		(16\pi G r)p_\perp(x,\lambda)&=&	\tfrac{5 (3 \lambda +1) (\lambda +x-1)^4}{8 \lambda ^3 (\lambda +1)^3}+\left(\tfrac{5 (3 \lambda +1) x}{4 \lambda ^3
			(\lambda +1)^3}-\tfrac{(3 \lambda +1) (5 \lambda +1)}{2 \lambda ^3 (\lambda +1)^3}\right) (\lambda+x-1)^3\cr
		&+&\left(\tfrac{3 \left(5 \lambda ^2+4 \lambda +1\right)}{2 \lambda ^2 (\lambda +1)^3}-\tfrac{3 (3 \lambda +1) (5
			\lambda +1) x}{4 \lambda ^3 (\lambda +1)^3}\right) (\lambda +x-1)^2 \\
&+&\tfrac{3 \left(5 \lambda ^2+4 \lambda +1\right) x
			(\lambda +x-1)}{2 \lambda ^2 (\lambda +1)^3} \;.
\nonumber
\end{eqnarray}


\subsection{Non-negativity of \texorpdfstring{$f(x,\lambda)$}{Lg}}

An important aspect of any regular solution is
that  $f(x,\lambda)$ remains non-negative throughout the spacetime.
It needs to be verified that this is indeed the case everywhere  inside
the translational layer.

Let us start here
by suitably rewriting   Eq.~(\ref{fx''}) for the second derivative of $f(x,\lambda)$,
\begin{eqnarray}
\label{fx''sim}
f''(x,\lambda)&=&\tfrac{4 \left(5 \lambda ^3+3 \lambda ^2-6 \lambda -2\right)+5 (3 \lambda +1) x^3-18 (3 \lambda +1) x^2-3 \left(5 \lambda ^3+3 \lambda ^2-21 \lambda -7\right) x}{4 \lambda ^3 (\lambda +1)^3}\;.
\end{eqnarray}

Clearly, the denominator is positive. As for the numerator,
this  is positive in the region $\;1-\lambda<x<x_1\;$ and negative in
the region $\;x_1<x<1+\lambda\;$,
where we have defined
\begin{equation}
 x_1\;=\;\frac{15 \lambda ^2-\sqrt{3} \sqrt{75 \lambda ^4+40 \lambda ^3+2 \lambda ^2+8 \lambda +3}+44 \lambda +13}{2 (15 \lambda +5)}\;<\;1+\lambda\;.
\end{equation}

It is then appropriate to  consider the  two regions separately:\\

1. \underline{$x\in (1-\lambda,x_1)$}:  In this region, $\;f''(x,\lambda)>0\;$,
so that  $\;f'(x,\lambda)\;$ is increasing and we can write $\;0=f'(1-\lambda,\lambda)<f'(x,\lambda)<f'(x_1,\lambda)\;$.  It follows that $\;f(x,\lambda)\;$ is increasing, meaning that  $\;0=f(1-\lambda,\lambda)<f(x,\lambda)<f(x_1,\lambda)\;$. In other words, $f(x,\lambda)$ is non-negative throughout this region.\\

2. \underline{$x\in (x_1,1+\lambda)$}:  In this portion of the layer, $\;f''(x,\lambda)<0\;$, and so $f'(x,\lambda)$ is decreasing. Thus,  $\;f'(x_1,\lambda)>f'(x,\lambda)>f'(1+\lambda,\lambda)=\frac{1}{(1+\lambda)^2}>0\;$, which means that $\;f(x,\lambda)\;$ is increasing and we then have  $\;0<f(x_1,\lambda)<f(x,\lambda)<f(1+\lambda,\lambda)\;$. It follows that  $f(x,\lambda)$ is positive everywhere in this region.

The conclusion is that $f(x,\lambda)$ is non-negative at all points in the transitional layer.


\subsection{Positivity of \texorpdfstring{$\rho(x,\lambda)$}{Lg}}

The physical validity of the solution also requires
the  energy density to be  non-negative throughout the spacetime.
We now check if  this is so inside the translational layer.

To this end, let us define $\;\tilde{\rho}(x,\lambda)=(8\pi G r^2)\rho(x,\lambda)\;$ and determine its second derivative by twice  differentiating
Eq.~(\ref{yikes}). After some simplification, this becomes
\begin{equation}
\label{rho''}
	\tilde{\rho}''(x,\lambda)\;=\;\tfrac{3 (x-1) \left(2 \left(5 \lambda ^3+3 \lambda ^2-6 \lambda -2\right)-5 (3 \lambda +1) x^2+10 (3 \lambda +1)
		x\right)}{2 \lambda ^3 (\lambda +1)^3} \;.
\end{equation}

Since the denominator of  $\tilde{\rho}''(x,\lambda)$ is manifestly positive,
our focus is on the numerator, which is
positive in the region $\;1<x<1+\lambda<\frac{\sqrt{\frac{10 \lambda ^3+6 \lambda ^2+3 \lambda +1}{3 \lambda +1}}}{\sqrt{5}}+1\;$ and negative in the region $\;1-\frac{\sqrt{\frac{10 \lambda ^3+6 \lambda ^2+3 \lambda +1}{3 \lambda +1}}}{\sqrt{5}}<1-\lambda<x<1\;$.  As in the previous subsection, we  consider
the two relevant sections separately:\\

1. \underline{$x\in (1,1+\lambda)$}:  In this region, $\;\tilde{\rho}''(x,\lambda)>0\;$, so that  $\;\tilde{\rho}'(x,\lambda)\;$ is increasing and, hence,
$\tilde{\rho}'(x,\lambda)<\tilde{\rho}'(1+\lambda,\lambda)=0\;$.
This in turns means that $\;\tilde{\rho}(x,\lambda)\;$ is decreasing,
which tells us that $\;\tilde{\rho}(x,\lambda)>\tilde{\rho}(1+\lambda,\lambda)=0\;$. Therefore,  $\tilde{\rho}(x,\lambda)$ is non-negative in this region.\\

2. \underline{$x\in (1-\lambda,1)$}:  In this half of the layer,   $\;\tilde{\rho}''(x,\lambda)<0\;$, and so  $\;\tilde{\rho}'(x,\lambda)\;$ is decreasing.
It then follows that $\;0=\tilde{\rho}'(1-\lambda,\lambda)>\tilde{\rho}'(x,\lambda)\;$. Then, since  $\;\tilde{\rho}(x,\lambda)\;$ is decreasing,
we have  $\;1=\tilde{\rho}(1-\lambda,\lambda)>\tilde{\rho}(x,\lambda)>\tilde{\rho}(1,\lambda)=\frac{1}{2}\;$. That is,  $\tilde{\rho}(x,\lambda)$
is  strictly  positive.

We can conclude
that  $\tilde{\rho}(x,\lambda)$ is non-negative throughout the transitional layer,
and likewise for the energy density $\rho(x,\lambda)$ as
these are related by a positive factor of $8\pi G r^2$.


\subsection{Maximal values of \texorpdfstring{$\rho'(x,\lambda)
\;{\rm and}\; p_{\perp}(x,\lambda)$}{Lg}}

We expect the magnitudes of the first derivative of the energy density and also the transverse
pressure to be very large, at least near the center of the transitional layer.
To see that this is indeed the case, it is simpler to look
at the closely related functions  $\;\tilde{\rho}(x,\lambda)=(8\pi G r^2)\rho(x,\lambda)\;$ and $\;\tilde{p}_\perp(x,\lambda)=(16\pi G r) p_\perp(x,\lambda)\;$.

The  extremal points for the first derivative of
  $\;\tilde{\rho}(x,\lambda)$  can be obtained
from  Eq.~(\ref{yikes}).  These are found to be at
\begin{equation}
	x_1\;=\;1\;,
\end{equation}
\begin{equation}\label{x2}
	x_2\;=\; \frac{-\sqrt{5} \sqrt{30 \lambda ^4+28 \lambda ^3+15 \lambda ^2+6 \lambda +1}+15 \lambda +5}{5 (3 \lambda +1)}\;,
\end{equation}
\begin{equation}\label{x3}
	x_3\;=\;\frac{\sqrt{5} \sqrt{30 \lambda ^4+28 \lambda ^3+15 \lambda ^2+6 \lambda +1}+15 \lambda +5}{5 (3 \lambda +1)}\;.
\end{equation}

Since $x_2$ and $x_3$ are not in the transitional layer, the relevant extremal point is $x_1$, which is at a  local minimum.
Thus, the maximally negative value of $\tilde{\rho}'(x,\lambda)$ is
\begin{equation}
	\tilde{\rho}_{\text{max}} '(x,\lambda)\;=\;\tilde{\rho}'(x_1,\lambda)\;=\;-\frac{3 \left(5 \lambda ^3+7 \lambda ^2+6 \lambda +2\right)}{8 \lambda  (\lambda +1)^3}\;.
\end{equation}
One can see that $\tilde{\rho}_{\text{max}} '(x,\lambda)$ is of the order $\frac{1}{\lambda}$ as pointed out in the main text.

The extremal points of $\tilde{p}_\perp(x,\lambda)$ can be found
by  analyzing  Eq.~(\ref{theperp}). One finds these to be at
\begin{equation}
	x_{1a}\;=\;1\;,
\end{equation}
\begin{equation}
	x_{2a}\;=\;\frac{-\sqrt{5} \sqrt{30 \lambda ^4+28 \lambda ^3+15 \lambda ^2+6 \lambda +1}+15 \lambda +5}{5 (3 \lambda +1)}\;,
\end{equation}
\begin{equation}
	x_{3a}\;=\;\frac{\sqrt{5} \sqrt{30 \lambda ^4+28 \lambda ^3+15 \lambda ^2+6 \lambda +1}+15 \lambda +5}{5 (3 \lambda +1)}\;.
\end{equation}

As the latter pair are not located in the transitional layer, the relevant extremal point is $x_{1a}$ and the maximal value of $\tilde{p}_\perp(x,\lambda)$ is
then
\begin{equation}
	\tilde{p}_\perp^{\text{max}}(x,\lambda)\;=\;\tilde{p}_\perp(x_{1a},\lambda)\;=\;\frac{3 \left(5 \lambda ^3+7 \lambda ^2+6 \lambda +2\right)}{8 \lambda  (\lambda +1)^3}\;.
\end{equation}

Let us pause here to note that  $\tilde{p}_\perp^{\text{max}}(x,\lambda)$
only has the single extremal point within the layer and is non-negative
at the endpoints. It must then be non-negative throughout the layer. The same obviously
applies to $p_{\perp}(x,\lambda)$, which  is also known
to vanish identically when
outside the layer. This along with the equation of state  $\;p_r+\rho=0\;$ confirms the validity of the null energy condition
throughout the spacetime.

One can now see that $\;\tilde{p}_\perp^{\text{max}}(x,\lambda)\;$ is of the same order $\frac{1}{\lambda}$ as found for $\tilde{\rho}_{\text{max}} '(x,\lambda)$.
That their respective maximal values  match,
could be deduced from an inspection
of the conservation equation in its reduced form,  Eq.~(\ref{E3}).

\subsection{Further consistency checks}

Given  the form of the line element in Eq.~(\ref{metric}), the
corresponding  Ricci scalar and Ricci tensor are expressible as
\begin{equation}
	R\;=\;-\frac{2(-1+f+2rf')}{f^2}-f''\;,
\end{equation}
\begin{equation}
	R^\mu_{\;\;\nu}\;=\;
	\begin{bmatrix}
		-\frac{f'}{r}-\frac{f''}{2}&0&0&0\\
		0&-\frac{f'}{r}-\frac{f''}{2}&0&0\\
		0&0& \frac{1}{r^2}(1-f-rf')&0\\
		0&0&0& \frac{1}{r^2}(1-f-rf')
	\end{bmatrix}\;.
\end{equation}

Let us also recall the basic form of the stress tensor for our special equation of state,
\begin{equation}
T^\mu_{\;\;\nu}\;=\;
\begin{bmatrix}
	-\rho&0&0&0\\
	0&-\rho&0&0\\
	0&0& p_\perp&0\\
	0&0&0&p_\perp
\end{bmatrix}\;.\label{energyTensor}
\end{equation}

We have used these relations and previous formalism to verify that
the solution in  the translational layer satisfies the Bianchi identity,
the conservation of the stress tensor and the Einstein equations.

\end{appendix}


\end{document}